\documentclass[journal]{IEEEtran}
\ifCLASSINFOpdf
\else
\fi
\usepackage{amssymb}
\usepackage{amsthm}
\usepackage{amsmath}
\usepackage{amsfonts}
\usepackage{graphicx}
\usepackage{graphics}
\usepackage{subfigure}
\usepackage{multirow}

\usepackage{tikz,pgfplots}
\usetikzlibrary{decorations.pathmorphing,decorations.pathreplacing,decorations.shapes}
\usetikzlibrary{shadows}
\usetikzlibrary{arrows}
\usetikzlibrary{shapes.arrows}
\usetikzlibrary{shapes.symbols}
\usepgflibrary{shapes.geometric}
\usetikzlibrary{mindmap,trees}

\begin{document}
%
\title{On Lattice Sequential Decoding for The Unconstrained AWGN Channel}
%
%
%

\author{Walid~Abediseid,~\IEEEmembership{Member,~IEEE} and
        Mohamed-Slim~Alouini,~\IEEEmembership{Fellow,~IEEE,}

\thanks{The authors are with the Electrical Engineering Program, Computer, Electrical and, Mathematical Sciences and Engineering (CEMSE) Division, King Abdullah University of Science and Technology (KAUST), Thuwal, Makkah Province, Saudi Arabia (e-mail: {walid.abediseid, slim.alouini}@kaust.edu.sa).}
\thanks{Manuscript received December 2012; revised February 5, 2013; accepted April 5, 2013.}}

%
%

\markboth{ACCEPTED for PUBLICATION in IEEE TRANSACTIONS ON COMMUNICATIONS,~Vol.~, No.~, APRIL~2013}%
{Shell \MakeLowercase{\textit{et al.}}: Bare Demo of IEEEtran.cls for Journals}
%



\maketitle

\begin{abstract}
In this paper, the performance limits and the computational complexity of the \textit{lattice sequential decoder} are analyzed for the unconstrained additive white Gaussian noise channel. The performance analysis available in the literature for such a channel has been studied only under the use of the minimum Euclidean distance decoder that is commonly referred to as the \textit{lattice decoder}. Lattice decoders based on solutions to the NP-hard closest vector problem are very complex to implement, and the search for low complexity receivers for the detection of lattice codes is considered a challenging problem. However, the low computational complexity advantage that sequential decoding promises, makes it an alternative solution to the lattice decoder. In this work, we characterize the performance and complexity tradeoff via the error exponent and the decoding complexity, respectively, of such a decoder as a function of the decoding parameter --- \textit{the bias term}. For the above channel, we derive the cut-off volume-to-noise ratio that is required to achieve a good error performance with low decoding complexity. 
\end{abstract}


%
\IEEEpeerreviewmaketitle

\section{Introduction}
\IEEEPARstart{T}{he} theory of \textit{lattices} --- a mathematical approach for representing infinite discrete points in Euclidean space~\cite{Conway}, has become a powerful tool to analyze many point-to-point and multi-terminal digital and wireless communication systems, particularly, communication systems that can be well described by the \textit{linear Gaussian vector channel} model. This is mainly due to the three facts about channel codes constructed using lattices: they have simple structure, their ability to achieve the fundamental limits (the capacity) of the channel, and most importantly, they can be decoded using efficient decoders called \textit{lattice decoders} \cite{deBuda}. Many researchers have studied the information-theoretic limits of lattice coding and decoding schemes for the linear Gaussian vector channel model \cite{deBuda}--\cite{Li}.

Poltyrev \cite{Polytrev} studied the problem of coding for the unconstrained additive white Gaussian noise (AWGN) channel where the channel input is an infinite lattice. In his setting, the notion of capacity becomes meaningless as infinite rates of transmission are possible. Therefore, another significant measurement was defined that characterizes the performance limits of such coding scheme when decoded using lattice decoders --- the \textit{normalized density} of the lattice or equivalently the information density rate of the lattice.

Based on a random lattice coding technique, Poltyrev showed that, using lattice decoding, the average probability of error can be upper bounded as
\begin{equation}\label{error_exponent}
P_{e,{\rm av}}(\mu_c)\leq e^{-mE_p(\mu_c)},
\end{equation}
where $m$ is the dimension of the lattice code, and $E_p(\mu_c)$ is called the \textit{Poltyrev error exponent} and is shown to be a non-zero, monotonically increasing, positive function for all $\mu_c>1$. The parameter $\mu_c$, which is called the \textit{volume-to-noise ratio}, to be defined in the sequel, is a quantity that is related to the density of the lattice. Hence, $\mu_c=1$ has the significance of capacity.

In \cite{Loe}, Loeliger proved that the above upper bound can be achieved using ensembles of \textit{linear} lattices --- constructed using linear codes over the ring of $p$-prime integer numbers, i.e., $\mathbb{Z}_p$, which is usually referred to as Construction~A \cite{Conway}. An important aspect of both Poltyrev's and Loeliger's proofs is based on an important theorem in number theory that is referred to as \textit{Minkowski-Hlawka} theorem~\cite{Hlawka},~\cite{Rogers}. 

It is clear from the above bound that large lattice codes would be required to approach capacity and therefore more practical decoding methods would be needed. It is well known that lattice decoders that are implemented using sphere decoding algorithms\footnote{Sphere decoding algorithms were originally implemented to decode signals transmitted via wireless fading channels \cite{JB}, particularly for the quasi-static multiple-input multiple-output wireless channels as an attempt to reduce the high computational complexity of the optimal maximum-likelihood decoder (see \cite{DGC}). The latter channel maybe described by the linear Gaussian vector channel model which allows the use of lattice coding, and lattice decoding to analyze the performance limits of such systems.} can be considered as a search in a \textit{tree} (see \cite{DGC}, \cite{JB} and references therein). Generally speaking, a sphere decoding algorithm explores the tree of all possible lattice points and uses a \textit{path metric} in order to discard paths corresponding to points outside the search sphere. Unfortunately, sphere decoding suffers from high computational complexity for low-to-moderate volume-to-noise ratios\footnote{The notion of ``signal-to-noise ratio'' is usually used for power-constrained channels where only a finite number of codewords or signals can be transmitted. Here, for infinite lattice codes, the notion of volume-to-noise ratio is used instead which will be introduced in the sequel.} and for large signal dimensionality in which low error probability is expected \cite{DGC}. As an alternative to sphere decoding algorithms, \textit{sequential decoders} comprise a set of efficient and powerful decoding techniques able to perform the tree search. These decoders can achieve \textit{near}-optimal performance without suffering the complexity of the sphere decoder for coding rates not too close to capacity\footnote{The work in \cite{MGDC} considered the application of lattice sequential decoders to various systems that can be described by the linear Gaussian vector channel model, such as the slowly-fading multiple-input multiple-output wireless channel, and the inter-symbol interference channel. In this work, it has been shown that near-sphere decoding performance can be achieved without suffering the high decoding complexity of the sphere decoder.} \cite{Jacobs},  \cite{Stack}.

The stack algorithm is a well known algorithm that is used to describe the operation of the sequential decoder \cite{Stack}. The algorithm was originally constructed as an alternative approach to the maximum-likelihood (ML) decoder for detecting convolutional codes transmitted via discrete memoryless channels. It has been shown in \cite{Jacobs} that as long as we operate below the cutoff rate, the decoder can achieve near-ML performance with low decoding complexity. 

For the lattice coded/decoded linear Gaussian vector channel model, there is a small body of work that discusses the performance and complexity tradeoff achieved by sequential decoding algorithms. Initial work on this topic was done by Tarokh \textit{et. al.} \cite{TVZ} where sequential decoding is used to decode lattice codes with finite trellis diagram. Shalvi \textit{et. al.} in \cite{OMN} has considered the use of sequential decoders to decode convolutional lattice codes. These power-limited (finite) lattice codes are generated using lattices combined with special lattice shaping techniques. The convolutional structure of such codes allows the use of the sequential decoders to achieve high data rates with low decoding complexity (this was mainly shown via simulation). However, all previous works lack a thorough theoretical analysis that can describe the systematic approach for tradeoff performance, complexity, and rate (or lattice density) achieved by sequential decoding of infinite lattice codes. 

This paper presents a complete performance analysis of the lattice sequential decoder in terms of the achievable error exponent. Moreover, the computational complexity of the decoder is determined via its complexity tail distribution where a new notion of the ``cut-off'' rate is defined. Both, the error performance and the decoder complexity, are derived as a function of the decoding parameter -- the \textit{bias term}. In order to fully characterize the performance of the decoder, we determine for the first time the error exponent achievable by lattice coding and sequential decoding applied to the unconstrained AWGN channel. We derive the error exponent as a function of the bias term which is critical for controlling the amount of computations required at the decoding stage. Achieving low decoding complexity requires increasing the value of the bias term. However, this is done at the expense of increasing erroneous detection. In this work, we follow the footsteps of Poltyrev and use the same definition of capacity for such a channel. We make use of lattice codes drawn from the ensemble of linear lattices, i.e., the Loeliger construction~\cite{Loe}.

We analyze in details the computational complexity tail distribution of the lattice sequential decoder. We show that there exists a cut-off volume-to-noise ratio that yields low decoding complexity which is also an increasing function of the bias term. We show that achieving low decoding complexity with good error performance comes at the expense of increasing the cut-off volume-to-noise ratio. Hence, lattice sequential decoders provide a systematic approach for tradeoff performance, complexity, and lattice density.

In contrast to most work in sequential decoding algorithms where the bias term is usually optimized to achieve a good performance-complexity tradeoff, we allow the bias term to vary freely and study the effect of this variation on the performance-complexity tradeoffs achieved by such decoders.

Throughout the paper, we use the following notation. The superscript $^\mathsf{T}$ denotes transpose. For a bounded region $\mathcal{R}\subset\mathbb{R}^m$, $V(\mathcal{R})$ denotes the volume of $\mathcal{R}$. We denote $\mathcal{S}_m(r)$ by the $m$-dimensional hypersphere of radius $r$ with $V(\mathcal{S}_m(r))=(\pi r^2)^{m/2}/\Gamma(m/2+1)$, where $\Gamma(m)=\int_0^\infty x^{m-1}e^{-x}\;dx$, is the Gamma function. Vectors are represented by bold lowercase letters, and matrices by bold uppercase letters where $\pmb{I}_m$ denotes the $m\times m$ identity matrix. The $l_2$-norm of a vector $\pmb{a}$ is denoted by $\|\pmb{a}\|$. The notation $\pmb{v}\sim\mathcal{N}(\pmb{\mu},\pmb{K})$ indicates that $\pmb{v}$ is a real Gaussian random vector with mean $\pmb{\mu}$ and covariance matrix $\pmb{K}$, and $\mathcal{E}\{\cdot\}$ represents the statistical average.

\section{Coding Without Restriction for the AWGN Channel}
\subsection{Lattice Properties}
A \textit{lattice} is a discrete pointset $\Lambda$ in a Euclidean space $\mathbb{R}^m$ that is closed under vector addition, i.e., any translate $\Lambda+\pmb{x}$ by a lattice point $\pmb{x}\in\Lambda$ is just $\Lambda$ again. Let $\{\pmb{g}_1,\pmb{g}_2,\cdots,\pmb{g}_m\}$ be a set of linearly independent vectors in $\mathbb{R}^m$. The set $\Lambda$ of all linear combinations $\pmb{x}=z_1\pmb{g}_1+z_2\pmb{g}_2+\cdots+z_m\pmb{g}_m$ with integer coefficients $z_i$ is a lattice, i.e.,
$$\Lambda=\{\pmb{x}=\pmb{Gz}:\pmb{z}\in\mathbb{Z}^m\},$$
where $\pmb{G}=[\pmb{g}_1,\pmb{g}_2,\cdots,\pmb{g}_m]$ is an $m\times m$ full-rank generator matrix. Thus, any lattice $\Lambda$ in $\mathbb{R}^m$ can be seen as a linear transformation of the integer lattice $\mathbb{Z}^m$. 

 Some properties associated with the lattice $\Lambda$ are of great importance for our analysis:
\begin{itemize}
\item The nearest neighbor quantizer $Q(\cdot)$ associated with $\Lambda$ is defined by
$$Q_\Lambda(\pmb{x})=\arg\min_{\pmb{\lambda}\in\Lambda}\|\pmb{\lambda}-\pmb{x}\|.$$

\item The Voronoi cell $\mathcal{V}(\pmb{\lambda})$ that corresponds to the lattice point $\pmb{\lambda}\in\Lambda$ is the set of points in $\mathbb{R}^m$ closest to $\pmb{x}$, i.e.,
$$\mathcal{V}(\pmb{\lambda})=\{\pmb{x}\in\mathbb{R}^m:Q_{\Lambda}(\pmb{x})=\pmb{\lambda}\}.$$
Voronoi cells associated with each lattice point $\pmb{\lambda}\in\Lambda$ are congruent and therefore can be considered as a shift of $\mathcal{V}(\pmb{0})$ by $\pmb{\lambda}$.

\item The volume of the Voronoi cell is given by
$$V(\pmb{G})\triangleq{\rm Vol}(\mathcal{V}(\pmb{0}))=\sqrt{\det\left(\pmb{G}^\mathsf{T}\pmb{G}\right)},$$
with the property that $V(a\pmb{G})=a^mV(\pmb{G})$ for any $a>0$.

\item The covering radius $r_{\rm cov}(\Lambda)$ is the radius of the smallest sphere centered at the origin that contains $\mathcal{V}(\pmb{0})$. The effective radius $r_{\rm eff}(\Lambda)$ is the radius of the sphere with volume equal to $V(\pmb{G})$. The packing radius $r_{\rm pack}(\Lambda)$ is the radius of the largest sphere centered at the origin inside the Voronoi cell $\mathcal{V}(\pmb{0})$ (see Fig.~\ref{fig:radius}).
\begin{figure}[htbp]
\centering
\includegraphics[width=2in]{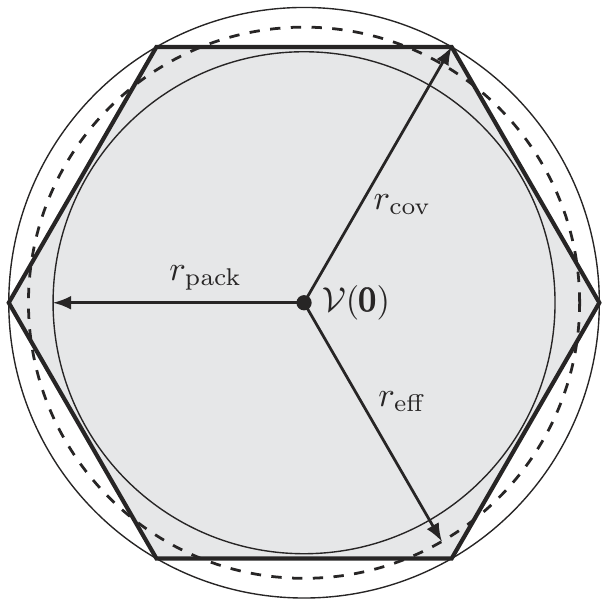}
\caption{The packing radius, the effective radius, and the covering radius of the hexagonal lattice.}
\label{fig:radius}
\end{figure}

\item A sequence of lattices $\{\Lambda_m\}$ of increasing dimension is good for covering \cite{EZ2} if $r_{\rm cov}(\Lambda_m)\rightarrow$ $r_{\rm eff}(\Lambda_m)$.

\item \textbf{Minkowski}-\textbf{Hlawka Theorem} \cite{Hlawka}: Let $f:\mathbb{R}^m\rightarrow\mathbb{R}$ be a Riemann integrable function of bounded support (i.e., $f(\pmb{x})=0$ if $\|\pmb{x}\|$ exceeds some bound). For any $\delta>0$, there exist ensembles $\pmb{\Lambda}=\{\Lambda\}$ of lattices with volume $V(\pmb{G})$ and dimension $m$ such that
\begin{equation}\label{MHT}
\mathcal{E}_{\pmb{\Lambda}}\left\{\sum\limits_{\pmb{x}\in\Lambda^*}f(\pmb{x})\right\}\leq (1+\delta){1\over V(\pmb{G})}\int_{\mathbb{R}^m}f(\pmb{x})\;d\pmb{x},
\end{equation}
where the expectation $\mathcal{E}_{\pmb{\Lambda}}$ is taken over the ensemble of random lattices, $\Lambda^*=\Lambda\backslash\{\pmb{0}\}$, and $\delta\rightarrow 0$ as $m\rightarrow \infty$. The above important theorem is sometimes regarded as a pre-Shannon result in information theory. In fact, the Mikowski-Hlawka theorem was originally used for packing lattices to solve the well known sphere-packing problem \cite{Rogers}.
\end{itemize}
\subsection{Poltyrev Error Exponent}
Suppose that an $m$-dimensional lattice point $\pmb{x}=\pmb{G}_c\pmb{z}\in\Lambda_c$ is to be transmitted through the unconstrained AWGN channel, where $\Lambda_c$ is an infinite lattice code with volume $V_c\stackrel{\Delta}{=}V(\pmb{G}_c)$, that is drawn from the ensemble of linear lattices using the Loeliger construction (see \cite{Loe} for more details about the construction). The received vector (output of the channel) in this case can be mathematically expressed as
\begin{equation}\label{AWGNy}
\pmb{y}=\pmb{x}+\pmb{w},
\end{equation}
where $\pmb{w}\sim\mathcal{N}(\pmb{0},\sigma^2\pmb{I}_m)$. Due to the unconstrained power condition on the lattice codewords (points), the optimum receiver that minimizes the probability of decoding error can be expressed as
\begin{equation}\label{opt0}
\hat{\pmb{z}}={\rm arg}\min_{\pmb{z}\in\mathbb{Z}^m}\|\pmb{y}-\pmb{G}_c\pmb{z}\|^2,
\end{equation}
which corresponds to searching over the whole lattice $\Lambda_c$ to find the closest point to the received vector $\pmb{y}$. This is referred to as lattice decoding.

As mentioned in the introduction, Poltyrev studied the problem of coding for the unconstrained AWGN channel with the input alphabet being the whole space $\mathbb{R}^m$. Since infinite power is possible, the notion of capacity becomes meaningless. Instead, the decoding error probability is measured against the normalized per dimension \textit{volume-to-noise ratio} (VNR), $\mu_c$, defined by
\begin{equation}
\mu_c\triangleq {V(\pmb{G}_c)^{2/m}\over2\pi e\sigma^2}={V_c^{2/m}\over 2\pi e\sigma^2},
\end{equation}
where $({V_c^{2/m}/ 2\pi e})\cdot m$ represents the asymptotic (in dimension $m$) squared radius of a sphere of volume $V_c$.

Poltyrev showed that the average probability of error (averaged over the ensemble of linear lattice codes $\Lambda_c$) is upper bounded by (\ref{error_exponent}) where 
\begin{equation}\label{E_p}
E_p(\mu_c)=\begin{cases}{1\over 2}\left[(\mu_c -1)-\log\mu_c\right], & 1<\mu_c\leq 2;\cr
                                            {1\over 2}\log\left(\displaystyle{e\mu_c\over 4}\right), & 2\leq \mu_c\leq 4;\cr
                                             \mu_c/8, &\mu_c\geq 4. \end{cases}
\end{equation}

From the above analysis, one notices that $\mu_c$ can be interpreted as the ratio of the squared radius of a spherical Voronoi cell to the variance of the noise. For small $\mu_c$, i.e., $\mu_c<1$, the spherical Voronoi cell has radius less than the standard deviation of the noise. In this case, reliable communication is not possible as error is highly likely to occur. As such, $\mu_c=1$ has the significance of capacity. 

Interestingly, Poltyrev showed that if only a \textit{finite} number of lattice points are to be transmitted as codewords with finite power constraint and transmission rate $R$, then rates $R$ up to ${1/2}\log(\text{SNR})$ are achievable, where $\text{SNR}\geq 0$ here represents the average signal-to-noise ratio of the channel\footnote{This can be simply done by intersecting the lattice code $\Lambda_c$ (possibly shifted by a vector $\pmb{u}_0$) with a shaping region $\mathcal{R}$ (a sphere or a Voronoi cell of another lattice), i.e., $\mathcal{C}=(\Lambda_c+\pmb{u}_0)\cap\mathcal{R}$. In this case, the transmission rate is given by $R={1\over m}\log_2[V(\mathcal{R)}/V_c]$, where $V(\mathcal{R})$ is the volume of the shaping region. If we define $mP_x={1\over|\mathcal{C}|}\sum_{\pmb{x}\in\mathcal{C}}\|\pmb{x}\|^2$ to be the average transmitted power, then one can show that $V(\mathcal{R})$ is asymptotically (as $m\rightarrow\infty$) given by $(2\pi eP_x)^{m/2}$. For reliable communication, we must have $V_c>(2\pi e\sigma^2)^{m/2}$. Therefore, rates $R$ up to ${1\over 2}\log(P_x/\sigma^2)={1\over 2}\log(\text{SNR})$ is achievable. }. For high SNRs (i.e., for $\text{SNR}\gg 1$), $1/2\log(\text{SNR})$ represents the capacity of the AWGN channel, denoted by $C$. Therefore, the same error probability bound given in (\ref{error_exponent}) and (\ref{E_p}) can be used (asymptotically) to characterize the performance of the power-limited lattice coded/decoded AWGN channel by letting (at high SNR) $\mu_c=2^{2[C-R]}$.    

Unfortunately, lattice decoders (usually implemented using sphere decoding algorithms) suffer from high computational complexity for low-to-moderate SNR and for large signal dimensionality in which low error probability is to be expected. As an alternative to lattice decoders, \textit{lattice sequential decoders} comprise a set of efficient and powerful decoding techniques that can achieve \textit{near}-optimal performance without suffering the complexity of the lattice decoder for coding rates not too close to capacity $C$. In fact, it is well known that sequential decoders can work well (with low decoding complexity) for rates below the cut-off rate $R_0$ which is only a factor of $4/e$ ($1.68$ dB) away from capacity $C$ at the high-SNR regime \cite{Gallager}. Therefore, for the unconstrained AWGN channel, $\mu_c=4/e$ has the significance of the cut-off rate. Here, we call this the \textit{cut-off VNR}, denoted by $\mu_0$.

\section{The Stack Sequential Decoder}
In this section, we briefly introduce the operation of the stack algorithm. This algorithm is an efficient tree search algorithm that attempts to find a ``best fit'' with the received noisy signal. Before we proceed with the description of such an algorithm, we shall discuss the metric measure for sequential decoding of lattice codes. It is basically based on the \textit{path metric} defined for conventional sequential decoders which is given by \cite{MGDC}
\begin{equation}
\label{FM}
\mathcal{M}(\pmb{z}_1^k)=\log\left({\Pr(\mathcal{H}(\pmb{z}_1^k))f(\pmb{y}_1^k|\mathcal{H}(\pmb{z}_1^k))\over f(\pmb{y}_1^k)}\right),
\end{equation}
where $\mathcal{H}(\pmb{z}_1^k)$ is the hypothesis that $\pmb{z}_1^k$ form the first $k$ symbols of the transmitted information sequence, and $f(\cdot)$ is the usual probability density function.

Recently, it has been shown that the search for the closest lattice point problem which corresponds to (\ref{opt0}) can be efficiently performed using sequential decoders based on the stack algorithm \cite{MGDC}. For our channel model, the path metric given by (\ref{FM}) can be shown to be simplified to (see Appendix A in~\cite{MGDC})
\begin{equation}
\label{metric}
\mathcal{M}(\pmb{z}_1^k)=bk-\|{\pmb{y}'}_1^k-\pmb{R}_{kk}\pmb{z}_1^k\|^2,
\end{equation}
where $\pmb{z}_1^k=[z_k,\cdots,z_2,z_1]^T$ denotes the last $k$ components of the integer vector $\pmb{z}$, $\pmb{R}_{kk}$ is the lower $k\times k$ matrix of $\pmb{R}$ that corresponds to the QR decomposition of the code matrix $\pmb{G}_c=\pmb{Q}\pmb{R}$, $\pmb{y}'=\pmb{Q}^\mathsf{T}\pmb{y}$, and $b$ is the bias term.

As in the conventional stack decoder \cite{Stack}, to determine a best fit (path), a value is assigned to each node in the tree. This value is called the metric which is given by (\ref{metric}). A flow chart for the operation of the stack decoder is shown in Fig.~2. As the decoder searches nodes, an ordered list of previously examined paths of different lengths is kept in storage. Each stack entry contains a path along with its metric. Each decoding step consists of extending the top (best) path in the stack. The determination of the best and next best nodes is simplified in the closest lattice point search problem by using the Schnnor-Euchner enumeration \cite{DGC} which generates nodes with metrics in ascending order given any node $\pmb{z}_1^k$. The decoding algorithm terminates when the top path in the stack reaches the end of the tree (refer to \cite{Stack} for more details about the algorithm).

\begin{figure}[ht!]
\begin{center}
\includegraphics[width=2in]{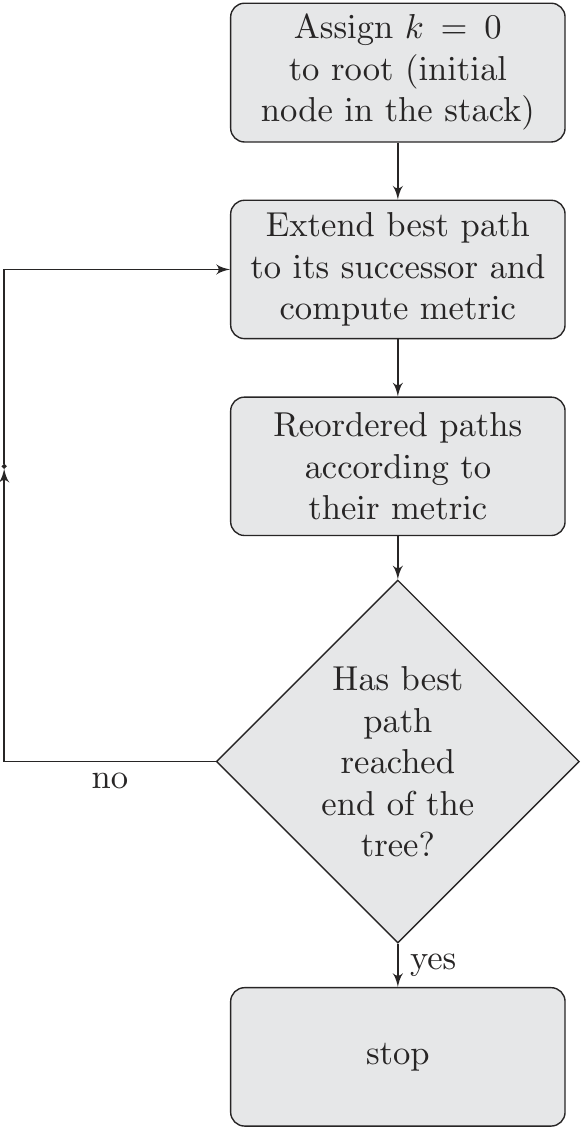}
\end{center}
\caption{Flow chart for stack decoding.}
\end{figure}

The main role of the bias term $b$ used in the algorithm is to control the amount of computations performed by the decoder. In this work, we define the computational complexity of the lattice sequential decoder as the \textit{total number of nodes visited by the decoder} during the search. Also, the bias term is responsible for the excellent performance-rate-complexity tradeoff achieved by such a decoding scheme. The role that the bias parameter plays will be discussed in detail in the subsequent sections.


\section{Performance Analysis: An Upper Bound}
As mentioned at the introduction, there has been no analysis devoted to sequential decoding applied to the lattice coded unconstrained AWGN channel. In this section, we analyze the performance limits of the stack sequential decoder when lattice coding is applied at the transmitter. We consider the unconstrained AWGN channel as defined by Poltyrev \cite{Polytrev}. Finding the exact error performance of such a decoder seems to be difficult. Therefore, we attempt to derive an upper bound on the sequential decoding error probability.

Define $P_e(b)$ as the probability that the sequential decoder makes an erroneous detection at a bias value $b$ (defined in (\ref{metric})). Now, due to lattice symmetry, one can assume that the all-zero lattice point $\pmb{0}$ is transmitted. Then, we have that
 \begin{equation}\label{UB202}
 \begin{split}
P_e(b)&=\Pr\left(\bigcup_{\pmb{x}\in\Lambda_c^*}\{\pmb{0}\text{ was decoded as }\pmb{x}\}\right)\cr
&\stackrel{(a)}{\leq}{\rm Pr}\left(\bigcup_{\pmb{z}\in\mathbb{Z}^m\backslash\{\pmb{0}\}}\{\mathcal{M}(\pmb{z})>\mathcal{M}_{\min}\}\right)\\
&\stackrel{(b)}{\leq}{\rm Pr}\left(\bigcup_{\pmb{x}\in\Lambda_c^*}\{\|\pmb{x}\|^2-2\pmb{x}^{\mathsf{T}}\pmb{w}<bm\}\right)\\
&={\rm Pr}\left(\bigcup_{\pmb{x}\in\Lambda_c^*}\left\{{2\pmb{x}^\mathsf{T}\pmb{w}}>{\|\pmb{x}\|^2}\left(1-{bm\over\|\pmb{x}\|^2}\right)\right\}\right),\cr
&\stackrel{(c)}{\leq}{\rm Pr}\left(\bigcup_{\pmb{x}\in\Lambda_c^*}\left\{{2\pmb{x}^\mathsf{T}\pmb{w}}>{\|\pmb{x}\|^2}\left(1-{bm\over d_{\min}^2(\Lambda_c)}\right)\right\}\right),
\end{split}
\end{equation}
where $\Lambda_c^*=\Lambda_c\backslash\{\pmb{0}\}$, $(a)$ is due to the fact that $\mathcal{M}(\pmb{z})>\mathcal{M}_{\min}$ is just a necessary condition for $\pmb{x}=\pmb{G}_c\pmb{z}$ to be decoded by the stack decoder, where $\mathcal{M}_{\min}=\min\{0,b-\|{\pmb{w}}_1^1\|^2,2b-\|{\pmb{w}}_1^2\|^2,\ldots,bm-\|{\pmb{w}}_1^m\|^2\}$ is the minimum metric that corresponds to the transmitted path, $(b)$ follows by noticing that $-(\mathcal{M}_{\min}+\|{\pmb{w}}\|^2)\leq 0$, and $(c)$ follows from the fact that $\|\pmb{x}\|\geq \min_{\pmb{x}\in\Lambda_c^*}\|\pmb{x}\|\stackrel{\Delta}{=}d_{\min}(\Lambda_c)$ --- the minimum Euclidean distance of the lattice.

It is clear from the above bound that the performance of the lattice sequential decoder depends critically on the shortest distance of the infinite lattice. Unfortunately, calculating the exact minimum distance of a lattice is NP-hard --- a problem that is referred to the shortest vector problem \cite{CLSP}. Moreover, finding the exact probability that appears in the RHS of (\ref{UB202}) for a particular lattice seems to be difficult. As such, we need to rely on a random technique to further upper bound the average error performance of the decoder. Before doing so, we need to ensure that the lattices in the ensemble are \textit{reasonably good} for channel coding. In order to do this, we need to expurgate the lattice ensemble that appears in (\ref{MHT}) appropriately such that the remaining lattices in the expurgated ensemble satisfy a lower bound on the packing radius of the lattice $r_{\rm pack}(\Lambda_c)$, or equivalently on the minimum Euclidean distance $d_{\min}(\Lambda_c)=2r_{\rm pack}(\Lambda_c)$. 

We recall the result in [6, Lemma 1] which states that most lattices in the random ensemble $\pmb{\Lambda}$ that satisfies the Minkowski-Hlawka theorem have good minimum Euclidean distance. In other words, for a lattice $\Lambda_c$ that is drawn from the random ensemble $\pmb{\Lambda}$ we have that for $0\leq \zeta<1$
\begin{equation}\label{d_bound}
\Pr(d_{\min}(\Lambda_c)>\zeta r_{\rm eff}(\Lambda_c))>1-\zeta^m,
\end{equation}
where $r_{\rm eft}(\Lambda_c)$ is the effective radius of $\Lambda_c$. Let $\pmb{\Lambda}_{\rm exp}$ be the expurgated lattice ensemble that satisfies (\ref{d_bound}), i.e.,
\begin{equation}
\label{expur}
 \pmb{\Lambda}_{\rm exp}(\zeta)=\{\Lambda_c\in\pmb{\Lambda}:d_{\min}(\Lambda_c)>\zeta r_{\rm eff}(\Lambda_c), 0\leq\zeta<1\}.
 \end{equation}

In this case, it is straight forward to show that for a given lattice $\Lambda_c\in\pmb{\Lambda}_{\rm exp}$, the conditional error probability (\ref{UB202}) can be further upper bounded by
 \begin{equation}\label{UB203}
 \begin{split}
P_e&(b|\Lambda_c)\leq\cr
&{\rm Pr}\left(\bigcup_{\pmb{x}\in\Lambda_c^*}\left\{{2\pmb{x}^\mathsf{T}\pmb{w}}>{\|\pmb{x}\|^2}\left(1-{bm\over \zeta^2 r^2_{\rm eff}(\Lambda_c)}\right)\right\}\Biggl|\Lambda_c\right).
\end{split}
\end{equation}
Averaging (\ref{UB203}) over the expurgated lattice ensemble, we get
 \begin{equation}\label{UB204}
 \begin{split}
\overline{P_e}(b)\leq\mathcal{E}_{\pmb{\Lambda_{\rm exp}}}\biggl\{{\rm Pr}\Biggl(\bigcup_{\pmb{x}\in\Lambda_c^*}\biggl\{{2\pmb{x}^\mathsf{T}\pmb{w}}&>{\|\pmb{x}\|^2}\bigl(1-\cr
&{bm\over \zeta^2 r^2_{\rm eff}(\Lambda_c)}\biggr)\biggr\}\Biggl|\Lambda_c\Biggr)\biggr\}.
\end{split}
\end{equation}

Now, in order to use the Minkowski-Hlawka theorem with expurgated ensemble we will need the following relation:
\begin{equation}
\mathcal{E}_{\pmb{\Lambda_{\rm exp}}}\left\{\pmb{X}\right\}\leq {1\over \Pr(d_{\min}(\Lambda_c)>\zeta r_{\rm eff}(\Lambda_c))}\mathcal{E}_{\pmb{\Lambda}}\left\{\pmb{X}\right\},
\end{equation}
where $\mathcal{E}_{\pmb{\Lambda}}\{\cdot\}$ is the expectation with respect to the ensemble in (\ref{MHT}), and $\pmb{X}$ is a nonnegative random variable. Therefore, we have that
 \begin{equation}\label{UB205}
 \begin{split}
\overline{P_e}(b)\leq{1\over 1-\zeta^m}\mathcal{E}_{\pmb{\Lambda}}\biggl\{{\rm Pr}\Biggl(\bigcup_{\pmb{x}\in\Lambda_c^*}\biggl\{{2\pmb{x}^\mathsf{T}\pmb{w}}&>{\|\pmb{x}\|^2}\biggl(1-\cr
&{bm\over \zeta^2 r^2_{\rm eff}(\Lambda_c)}\biggr)\biggr\}\biggl|\Lambda_c\Biggr)\biggr\}.
\end{split}
\end{equation}
 
As $m\rightarrow\infty$ we have 
$$r^2_{\rm eff}(\Lambda_c)={\Gamma\left({m\over 2}+1\right)^{2/m}\over \pi}V_c^{2/m} \sim{V^{2/m}_c\over 2\pi e}\cdot m=\mu_c\sigma^2m,$$
where $\mu_c$ is the VNR, and $\sigma^2$ is the noise variance. Note that, asymptotically, as $m\rightarrow\infty$, we may let $\zeta$ approaches $1$ as close as desired. Therefore, the average probability of decoding error can be asymptotically upper bounded by
 \begin{equation}\label{UB_final}
 \begin{split}
\overline{P_e}(b,\mu_c)&\leq \mathcal{E}_{\pmb{\Lambda}}\left\{{\rm Pr}\left(\bigcup_{\pmb{x}\in\Lambda_c^*}\left\{{2\pmb{x}^\mathsf{T}\pmb{w}}>{\|\pmb{x}\|^2}\left(1-{b/\sigma^2\over \mu_c}\right)\right\}\right)\right\}\cr
&=\mathcal{E}_{\pmb{\Lambda}}\left\{{\rm Pr}\left(\bigcup_{\pmb{x}\in\Lambda_c^*}\left\{{2{\pmb{x}}^\mathsf{T}\tilde{\pmb{w}}}>{\|\pmb{x}\|^2}\right\}\right)\right\},
\end{split}
\end{equation}
where 
$$\tilde{\pmb{w}}=\left(1-{b_{ n}\over \mu_c}\right)^{-1}\pmb{w},$$
is a zero-mean Gaussian random vector with elements that are independent, identically, distributed random variables with variance $\tilde{\sigma}^2=(1-b_n/\mu_c)^{-2}\sigma^2$, and $b_n=b/\sigma^2$ is defined as the normalized bias with respect to the noise variance. It must be noted that the above bound is only valid for all values of $b_n$ such that $1-b_n/\mu_c>0$, or equivalently for all values of $0\leq b_n<\mu_c$.

Interestingly, the upper bound (\ref{UB_final}) corresponds to the probability of decoding error of a received signal $\pmb{y}=\pmb{x}+\tilde{\pmb{w}}$ decoded using the conventional lattice decoder. Therefore, one may observe that the sub-optimality of the sequential decoder can be viewed as a source of channel noise amplification.

Following the footsteps of Poltyrev, the average probability of error can be shown to be upper bounded by
\begin{equation}
\overline{P_e}(b,\mu_c)\leq e^{-m E_b(\mu_c)},
\end{equation}
where 
\begin{equation}
E_b(\mu_c)=E_p(\tilde{\mu}_c)=\begin{cases}0, & \tilde{\mu}_c\leq 1;\cr 
{1\over 2}\left[(\tilde{\mu}_c -1)-\log\tilde{\mu}_c\right], & 1<\tilde{\mu}_c\leq 2;\cr
                                            {1\over 2}\log\left(\displaystyle{e\tilde{\mu}_c\over 4}\right), & 2\leq \tilde{\mu}_c\leq 4;\cr
                                             \tilde{\mu}_c/8, &\tilde{\mu}_c\geq 4. \end{cases}
\end{equation}
where 
 \begin{equation}\label{mu_c_p}
\tilde{\mu}_c\triangleq {{V(\pmb{G}_c)^{2/m}/ 2\pi e}\over\tilde{\sigma}^2}=\mu_c\left(1-{b_n\over \mu_c}\right)^{2}.
\end{equation}

Hence, for sufficiently large $m$, there exists at least a lattice $\Lambda'_c$ in the expurgated code ensemble with error probability satisfying 
\begin{equation}\label{avg_prob}
P_e(b,\mu_c,\Lambda'_c)\leq e^{-m E_b(\mu_c)}.
\end{equation}

Now, the following important remarks can be made about the above result:

\begin{itemize}
\item \textit{\textbf{Fixed Bias}}: In this case, the bias term $b$ is fixed and chosen independent of the VNR $\mu_c$. Note that as $\mu_c$ gets large ($\mu_c\gg 1$) , one may approximate $\tilde{\mu}_c$ in (\ref{mu_c_p}) as
\begin{equation}
\tilde{\mu}_c=\mu_c\left(1-{b_n\over \mu_c}\right)^{2}\approx\mu_c\left(1-2{b_n\over\mu_c}\right)=\mu_c-2b_n.
\end{equation}

Therefore, the above analysis indicates that fixing the bias term causes a right-shift to the error probability curve (i.e., a reduction in the coding gain). This can be realized from the value of the error exponent for large $\mu_c$ which is given by 
\begin{equation}\label{exp_approx}
E_b(\mu_c)={\tilde{\mu}_c\over 8}={\mu_c\over 8}-{b_n\over 4}=E_p(\mu_c)-{b_n\over 4},
\end{equation}
where $E_p(\mu_c)$ is the Poltyrev error exponent achieved by the lattice decoder which is defined in (\ref{E_p}). Substituting (\ref{exp_approx}) into (\ref{avg_prob}) we get
\begin{equation}\label{17}
\begin{split}
P_e(b,\mu_c)\leq e^{-m E_b(\mu_c)}&=e^{-m [E_p(\mu_c)-b_n/4]}\cr
&=\alpha e^{-m E_p(\mu_c)},
\end{split}
\end{equation}
where $\alpha=e^{mb_n/4}$. The constant $\alpha$ describes the behavior of the error probability of the sequential decoder for a fixed bias term. Increasing the bias term results in performance reduction compared to the one achieved by the lattice decoder. This reduction is represented by a right-shift to the error probability curve\footnote{It must be noted that, although the bound (\ref{avg_prob}) shows that the shift is $\alpha=e^{mb_n/4}$, the exact amount of right-shift is less than $\alpha$ as will be shown by the simulation results in Section VI.}, as will be shown in the sequel.
\item \textit{\textbf{Variable Bias}}: Now, let the normalized bias $b_n$ to scale linearly with the VNR $\mu_c$ as $b_n=(1-\sqrt{\delta})\mu_c$ where $0<\delta\leq 1$, then the error exponent in this case can be expressed as
\begin{equation}\label{188}
E_b(\mu_c)=\begin{cases}0, & \mu_c\leq 1/\delta;\cr 
{1\over 2}\left[(\delta\mu_c -1)-\log\delta\mu_c\right], & 1/\delta<\mu_c\leq 2/\delta;\cr
                                            {1\over 2}\log\left(\displaystyle{e\delta\mu_c\over 4}\right), & 2/\delta\leq \mu_c\leq 4/\delta;\cr
                                             \delta\mu_c/8, &\mu_c\geq 4/\delta. \end{cases}
\end{equation}

It is clear from the above analysis that if $\delta\rightarrow1$ (or $b_n\rightarrow0$) then the performance of the sequential decoder approaches the performance of the lattice decoder. On the other extreme, if $\delta\rightarrow0$ (or $b_n\rightarrow\mu_c$) then reliable communication may not be possible under lattice sequential decoding. Fig.~3 shows the error exponent achieved by the lattice sequential decoder for the case of the variable bias term described above. It is clear from Fig.~3 that for high VNR $\mu_c$, the effect of varying $\delta$ occurs as a change in the \textit{slope} of the error exponent curve, where at high VNR we have $E_b(\mu_c)=\delta\mu_c/8$. Moreover, the maximum achievable VNR under sequential decoding with normalized bias $b_n=(1-\sqrt{\delta})\mu_c$ is given by $1/\delta$. Therefore, for $\delta\neq 1$, reliable communication may not be possible at VNR close to capacity ($\mu_c=1$).
\end{itemize}
\begin{figure}[ht!]
\center
\includegraphics[width=3.5in]{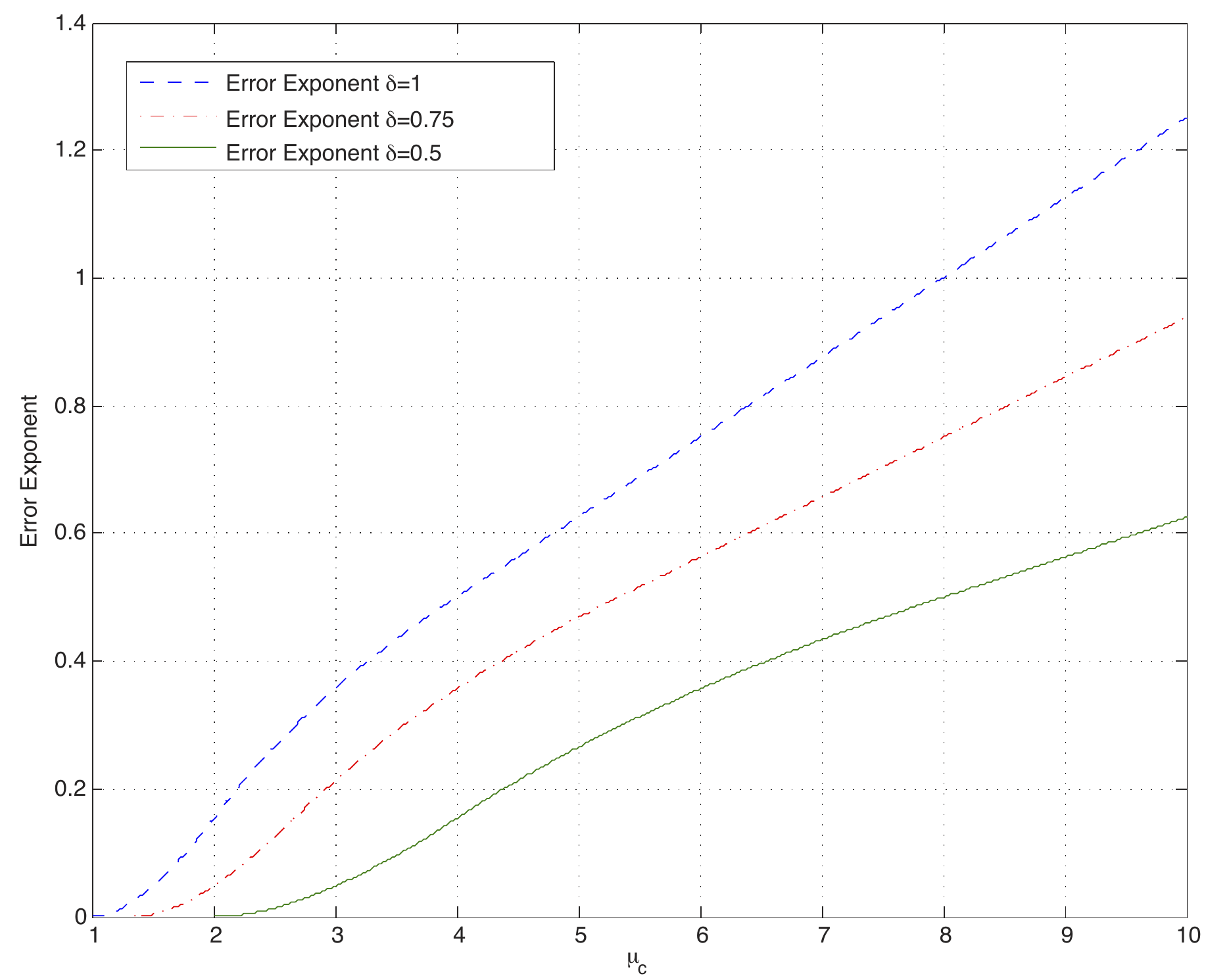}
\caption{The achievable error exponent of the lattice sequential decoder when the normalized bias term $b_n=(1-\sqrt{\delta})\mu_c$ for $\delta=0.5, 0.75,\text{ and } 1$.}
\end{figure}

The main result that we draw from the above discussion is that, increasing the bias term lower the decoding error probability. However, the loss in the error performance achieved by any sub-optimal decoder is usually compensated by some improvements in the decoding complexity. This fact will be demonstrated next where we consider the computational complexity analysis of the sequential decoder for the unconstrained AWGN channel.

\section{The ``cut-off'' Volume-to-Noise Ratio: An Upper Bound on the Complexity Distribution}
The main use of the sequential decoder is to achieve a low decoding complexity compared to the very complex lattice decoder. As in conventional sequential decoder, one need to back-off from capacity to achieve such improvements. For convolutional codes detected using sequential decoders, a cut-off rate has been defined for such decoding schemes. The cut-off rate $R_{0}$ is the rate for which the transmitter should not exceed if one needs to expect a low decoding complexity. If $R>R_{0}$, then the complexity of the sequential decoder increases exponentially with the constraint length of the code~\cite{Jacobs}. For the power-constrained AWGN channel, the cut-off rate $R_0$ is $4/e$ ($1.68$ dB) away from capacity at high SNR~\cite{Gallager}. In this section, we study in details the complexity of the sequential decoder for the unconstrained AWGN channel. 

Due to the random nature of the channel noise, the computational complexity of the lattice sequential decoder is also random. Therefore, it would be more appropriate to study the complexity behavior of such a decoder via its complexity tail distribution defined as $\Pr(\mathcal{N}(\Lambda_c)\geq L)$, where $\mathcal{N}(\Lambda_c)$ is defined as  the \textit{total number of nodes in the tree that have been visited by the decoder during the search} for a given lattice $\Lambda_c$, and $L$ is the distribution parameter. 

Similar to the power-limited AWGN channel, we define a ``cut-off'' VNR $\mu_0$ for the unconstrained AWGN channel to be the value of $\mu_c$ for which both low decoding complexity and low decoding error probability are possible. 

To simplify the analysis, we start by bounding the total number of computations $\mathcal{N}(\Lambda_c)$ from above as follows. First, one should note that all nodes in the tree that have been visited by the sequential decoder must have partial path metrics $\mathcal{M}(\pmb{z}_1^k)$ that exceed the minimum metric $\mathcal{M}_{\min}$ which corresponds to the decoded path. Let $\phi(\pmb{z}_1^k)$ be the indicator function defined by
\begin{equation}\label{00}
\phi(\pmb{z}_1^k)=\begin{cases} 1, &{\mathcal{M}(\pmb{z}_1^k)\geq \mathcal{M}_{\min};}\cr
                                                          0, &\text{otherwise,}\end{cases}
\end{equation}   
Since $\mathcal{M}(\pmb{z}_1^k)\geq \mathcal{M}_{\min}$ is a sufficient condition for the node to be visited by the decoder, then $\mathcal{N}(\Lambda_c)$ may be upper bounded by
\begin{equation}\label{complex_num}
\mathcal{N}(\Lambda_c)\leq\sum_{k=1}^{m}\sum_{\pmb{z}_1^k\in\mathbb{Z}^k}\phi(\pmb{z}_1^k).
\end{equation}

Also, the complexity tail distribution can be upper bounded as
\begin{align}\label{UBb1}
\Pr(\mathcal{N}(\Lambda_c)\geq L)\leq &\Pr(\mathcal{N}(\Lambda_c)\geq L,\|\pmb{w}\|^2\leq \sigma^2m)+\cr
&\quad\Pr(\|\pmb{w}\|^2>\sigma^2m),
\end{align}
where the above upper bound is derived using the well known separation of the typical noise events from the non-typical ones~\cite{Gallager}. Next, we would like to upper bound the first term in the RHS of (\ref{UBb1}). 

Given $\|\pmb{w}\|^2\leq \sigma^2m$, and by noticing that $-(\mathcal{M}_{\min}+\|\pmb{w}\|^2)\leq 0$, we obtain
\begin{equation}
\sum_{\pmb{z}_1^k\in\mathbb{Z}^k}\phi(\pmb{z}_1^k)\leq\sum_{\pmb{z}_1^k\in\mathbb{Z}^k}\phi^{'}(\pmb{z}_1^k),
\label{aa}
\end{equation}
where $\phi(\pmb{z}_1^k)$ is the indicator function defined in (\ref{00}), and
\begin{equation}\label{C111}
\phi^{'}(\pmb{z}_1^k)=\begin{cases} 1, &\text{if $\|{\pmb{w}'}_1^k-\pmb{R}_{kk}\pmb{z}_1^k\|^2\leq bk+\sigma^2m$;}\cr
                                                          0, &\text{otherwise.}\end{cases}
                                                          \end{equation}
where ${\pmb{w}'}_1^k$ is the last $k$ components of $\pmb{w}'=\pmb{Q}^\mathsf{T}\pmb{w}$. Now, let
$$
\phi^{''}_k(\pmb{z})=\begin{cases} S_k, &\text{if $\|\pmb{w}'-\pmb{R}\pmb{z}\|^2\leq bm-\mathcal{M}_{\min}$;}\cr
                                                          0, &\text{otherwise,}\end{cases}
$$
where
\begin{equation}\label{Sk}
S_k=\sum_{\pmb{z}_1^k\in\mathbb{Z}^k}\phi^{'}(\pmb{z}_1^k),
\end{equation}
then it can be shown that
$$\mathcal{N}(\Lambda_c)\leq\sum\limits_{k=1}^m\sum_{\pmb{z}\in\mathbb{Z}^m}\phi^{''}_k(\pmb{z})\leq\sum\limits_{k=1}^m\sum_{\pmb{x}\in\Lambda_c}\tilde{\phi}_k(\pmb{x}),$$
where
$$\tilde{\phi}_k(\pmb{x})=\begin{cases} S_k, &\text{if $\|\pmb{x}\|^2-2(\pmb{x})^\mathsf{T}\pmb{w}\leq bm$;}\cr
                                                          0, &\text{otherwise.}\end{cases}
                                                         $$
                                                        
Interestingly, the sum that appears in (\ref{Sk}) represents the number of partial integer lattice points $\pmb{z}_1^k\in\mathbb{Z}^k$ that are located inside a sphere of squared radius $bk+\sigma^2m$ centered at the received signal ($\pmb{y}=\pmb{w}$ in our case). One can approximate $S_k$ by the ratio of the volume of the $k$-dimensional sphere of squared radius $bk+\sigma^2m$ to the volume of the Voronoi cell of the lattice $\Lambda_k$ that corresponds to $\pmb{R}_{kk}$, denoted by $V(\pmb{R}_{kk})$, (see \cite{Rogers} for more 
details), i.e.,
\begin{equation}\label{lemma}
S_k\approx{\mathcal{S}_k(\sqrt{bk+\sigma^2m})\over V(\pmb{R}_{kk})}={{(\pi)}^{k/2}\over \Gamma(k/2+1)}{[bk+\sigma^2m]^{k/2}\over \det(\pmb{R}_{kk}^\mathsf{T}\pmb{R}_{kk})^{1/2}}.
\end{equation}

For a given lattice $\Lambda_c$, we have
\begin{equation}
\begin{split}
\Pr(\mathcal{N}&(\Lambda_c)\geq L,\|\pmb{w}\|^2\leq \sigma^2m|\Lambda_c)\cr
&\leq\Pr(\tilde{\mathcal{N}}(\Lambda_c)\geq L-m,\|\pmb{w}\|^2\leq \sigma^2m|\Lambda_c)\cr
&\leq{\mathcal{E}_{\pmb{w}}\{\tilde{\mathcal{N}}(\Lambda_c)|\Lambda_c,\|\pmb{w}\|^2\leq \sigma^2m\}\over L-m}, \quad \text{for $L>m$,}
\end{split}
\end{equation}
where the last inequality follows from using Markov inequality, and $\tilde{\mathcal{N}}(\Lambda_c)$ is defined as
$$\tilde{\mathcal{N}}(\Lambda_c)=\sum_{k=1}^{m}\sum_{\pmb{z}_1^k\in\mathbb{Z}^k\backslash\{\pmb{0}\}}\phi(\pmb{z}_1^k),$$
since we have assumed that the all-zero lattice point was transmitted.

The conditional average of $\tilde{\mathcal{N}}(\Lambda_c)$ with respect to the noise can be further upper bounded as
\begin{equation}\label{NUB1}
\begin{split}
\mathcal{E}_{\pmb{w}}\{\tilde{\mathcal{N}}(\Lambda_c)&|\Lambda_c,\|\pmb{w}\|^2\leq  \sigma^2m\}\cr
&\leq \sum\limits_{k=1}^m S_k\sum\limits_{\pmb{x}\in\Lambda_c^*}\Pr(\|\pmb{x}\|^2-2(\pmb{x})^{\mathsf{T}}\pmb{w}<bm).
\end{split}
\end{equation}
Therefore, we have
\begin{align}\label{EE1}
\Pr(&\mathcal{N}(\Lambda_c)\geq L,\|\pmb{w}\|^2\leq \sigma^2m|\Lambda_c)\cr
&\leq{\sum_{k=1}^m S_k\over L-m}\sum\limits_{\pmb{x}\in\Lambda_c^*}\Pr(2(\pmb{x})^{\mathsf{T}}\pmb{w}>\|\pmb{x}\|^2-bm).
\end{align}

Now, for $L=m+\sum_{k=1}^m S_k$, we have that
\begin{equation}\label{27}
\begin{split}
&\Pr(\mathcal{N}(\Lambda_c)\geq L|\Lambda_c)\cr
&\quad\leq\sum\limits_{\pmb{x}\in\Lambda_c^*}\Pr(2(\pmb{x})^{\mathsf{T}}\pmb{w}>\|\pmb{x}\|^2-bm)+\Pr(\|\pmb{w}\|^2>\sigma^2m)\\
&\quad\leq \sum\limits_{\pmb{x}\in\Lambda_c^*}\Pr\left(2(\pmb{x})^{\mathsf{T}}\pmb{w}>\|\pmb{x}\|^2\left(1-{bm\over d_{\min}^2(\Lambda_c)}\right)\right)\cr
&\hspace{5.35cm}+\Pr(\|\pmb{w}\|^2>\sigma^2m).
\end{split}
\end{equation}

At this point, it is worth mentioning that the first term in the RHS of (\ref{27}) represents the \textit{union bound} of the sequential decoding error probability in (\ref{UB202}). Therefore, as will be shown in the sequel, there exists a minimum VNR $\mu_c$, defined as the cut-off VNR $\mu_0$, such that for all $\mu_c>\mu_0$ low decoding error probability and low decoding complexity can be achieved simultaneously.

Similar to the error probability analysis, assuming $\Lambda_c$ is drawn from the expurgated lattice ensemble $\pmb{\Lambda}_{\rm exp}$ defined in~(\ref{expur}), we have that 
\begin{equation*}
\begin{split}
&\Pr\left(2(\pmb{x})^{\mathsf{T}}\pmb{w}>\|\pmb{x}\|^2\left(1-{bm\over d_{\min}^2(\Lambda_c)}\right)\right)\cr
&\hspace{2cm}\leq\Pr\left(2(\pmb{x})^{\mathsf{T}}\pmb{w}>\|\pmb{x}\|^2\left(1-{bm\over \zeta^2r_{\rm eff}^2(\Lambda_c)}\right)\right).
\end{split}
\end{equation*}
where $0<\zeta<1$. Using Chernoff bound, as $m\rightarrow\infty$
\begin{equation*}\label{UB5}
\Pr\left(2(\pmb{x})^{\mathsf{T}}\pmb{w}>\|\pmb{x}\|^2\left(1-{bm\over \zeta^2r_{\rm eff}^2(\Lambda_c)}\right)\right)\leq e^{-\|\pmb{x}\|^2/8\tilde{\sigma}^2},
\end{equation*}
where $\tilde{\sigma}^2=(1-b_n/\mu_c)^{-2}\mu_c$, and $b_n=b/\sigma^2$. Therefore, one may asymptotically upper bound the first term in the RHS of~(\ref{27}) as
\begin{equation}\label{F_B}
\sum\limits_{\pmb{x}\in\Lambda_c^*}\Pr(2(\pmb{x})^{\mathsf{T}}\pmb{w}>\|\pmb{x}\|^2-bm)\leq\sum\limits_{\pmb{x}\in\Lambda_c^{*}}e^{-\|\pmb{x}\|^2/8\tilde{\sigma}^2}.
\end{equation}

Substituting (\ref{F_B}) in (\ref{27}), and taking the expectation of (\ref{27}) over the ensemble of expurgated lattices, we obtain
\begin{equation}\label{Mink11}
\begin{split}
\overline{\Pr(\mathcal{N}(\Lambda_c)\geq L)}&=\mathcal{E}_{\pmb{\Lambda}_{\rm exp}}\{\Pr(\mathcal{N}(\Lambda_c)\geq L|\Lambda_c)\}\cr
                                &\leq{1\over V_c}\int\limits_{\mathbb{R}^m}e^{-\|\pmb{x}\|^2/8\tilde{\sigma}^2}\;d\pmb{x}+\Pr(\|\pmb{w}\|^2>\sigma^2m).
\end{split}
\end{equation}
Evaluating the integral in the above upper bound we get
\begin{equation}\label{f}
{1\over V_c}\int\limits_{\mathbb{R}^m}e^{-\|\pmb{x}\|^2/8\tilde{\sigma}^2}\;d\pmb{x}={(8\pi \tilde{\sigma}^2)^{m/2}\over V_c}=\left({4/e\over \tilde{\mu}_c}\right)^m,
\end{equation}
where $\tilde{\mu}_c=\mu_c\left(1-\displaystyle{b_n/ \mu_c}\right)^2$. Therefore, for large $m$, we can further upper bound (\ref{Mink11}) for all $0\leq b_n\leq\mu_c$ as
\begin{equation}\label{29}
\overline{\Pr(\mathcal{N}(\Lambda_c)\geq L)}\leq  \left({4/e\over \tilde{\mu}_c}\right)^m+\Pr(\|\pmb{w}\|^2>\sigma^2m).
\end{equation}

Hence, for sufficiently large $m$, there exists at least a lattice $\Lambda'_c$ in the ensemble with complexity tail distribution satisfying~(\ref{29}) for all values of $L\geq m+\sum_{k=1}^m S_k$. It follows from standard typicality arguments that for any $\epsilon>0$ there exists $m_0$ such that for all $m>m_0$
$$\Pr(\|\pmb{w}\|^2>\sigma^2m)<\epsilon/2.$$ 
The first term in the upper bound (\ref{29}) can be made smaller than $\epsilon/2$ for sufficiently large $m$, i.e.,
$$\Pr(\mathcal{N}(\Lambda'_c)\geq L)\leq \epsilon,\quad m\rightarrow\infty, $$
if $\tilde{\mu}_c>4/e$. This result indicates that large computational complexity may be avoided, while maintaining good error performance, at $\mu_c$ above the cut-off VNR $\mu_0$ which is given by the roots of the following equation
$$\mu_0\left(1-{b_n\over \mu_0}\right)^2={4\over e}.$$
Note that when $b_n=0$ we have $\mu_0=4/e$. Under this constraint, solving the above equation for $\mu_0$ we get
\begin{equation}\label{cut-off}
\mu_0=\left(b_n+{2\over e}\left[1+\sqrt{b_ne+1}\right]\right).
\end{equation}

It is interesting to note that as $b\rightarrow 0$ (the value of the bias that achieves close to lattice decoding performance) we have $\mu_0\rightarrow 4/e$, where $4/e$ represents the gap between the cut-off rate and the capacity of the power-constraint AWGN channel~\cite{Forney2},~\cite{Gallager}. Since the union bound in (\ref{27}) provides a good estimate to the decoding error probability at high VNR (i.e., for $\mu_c$ greater than the cut-off VNR $\mu_0$) (see \cite{Forney2}), achieving a good error performance for large values of $b$, where low decoding complexity is expected, comes at the expense of increasing the VNR (or equivalently reducing the coding rate for the case of finite lattice codes). 

The analysis above indicates that the total number of computations that is required by the decoder to decode a message while achieving low error probability can be approximated by
\begin{equation}\label{L1}
L\approx m + \sum\limits_{k=1}^m{{(\pi)}^{k/2}\over \Gamma(k/2+1)}{[bk+\sigma^2m]^{k/2}\over (\det(\pmb{R}_{kk}^\mathsf{T}\pmb{R}_{kk})^{1/2})}.
\end{equation}

In order to see how the complexity is affected by the channel and the decoder parameters, we express the unconstrained AWGN channel by the vector model $\pmb{y}=\sqrt{\mu_c}\pmb{x}+\pmb{w}$, where $\pmb{x}$ is the transmitted lattice point that is selected randomly from a lattice $\Lambda_c$ with generator matrix $\pmb{G}_c=\pmb{QR}$, $\mu_c$ is the VNR, and $\pmb{w}\sim\mathcal{N}(\pmb{0},\pmb{I})$. The volume of the Voronoi cell of $\Lambda_c$ is selected such that the VNR at the output of the channel is $\mu_c$. In this case we have $V_c=(2\pi e)^{m/2}$. As a result, we may express $L$ as 
\begin{equation}\label{L2}
L\approx m + \sum\limits_{k=1}^m{{(\pi)}^{k/2}\over \Gamma(k/2+1)}{[b_nk+m]^{k/2}\over \mu_c^{k/2}\det(\pmb{R}_{kk}^\mathsf{T}\pmb{R}_{kk})^{1/2}}.
\end{equation}

It is clear from the above equation that as $\mu_c\rightarrow\infty$, we have $L\rightarrow m$. Therefore, regardless the value of the bias term chosen at the decoding stage, the complexity of the decoder is approximately linear in the code dimension when the VNR is very large. This fact is also verified experimentally as will be shown in the sequel. 

In fact, as expected, the total number of computations performed by the decoder at any VNR $\mu_c$ beyond $\mu_0$ is bounded from above. This can be seen by substituting (\ref{cut-off}) in (\ref{L2}). Then, one may upper bound the total number of computations performed by the decoder as
\begin{equation}
\begin{split}
L&\leq m+\sum\limits_{k=1}^m {{(\pi)}^{k/2}\over \Gamma(k/2+1)}{[b_nk+m]^{k/2}\over \mu_0^{k/2}\det(\pmb{R}_{kk}^\mathsf{T}\pmb{R}_{kk})^{1/2}}\cr
&\approx m+\sum\limits_{k=1}^m {{(\pi k)}^{k/2}\over \Gamma(k/2+1)\det(\pmb{R}_{kk}^\mathsf{T}\pmb{R}_{kk})^{1/2}}, \quad b_n\rightarrow\infty.
\end{split}
\end{equation}

It is clear from the above bound that as $b$ increases (or as $b_n\rightarrow \infty$), the computational complexity scales almost linearly with the code dimension $m$. The simulation results (introduced next) agree with the above analysis.

In conclusion, lattice sequential decoders allow for a systematic approach for trading off performance, VNR, and complexity. For a fixed VNR, increasing the bias term allows to achieve low decoding complexity but at the expense of poor performance. In order to improve the performance without affecting the complexity, one need to increase the VNR $\mu_c$ or equivalently to increase the lattice density, to recover the performance loss. 

\section{Simulation Results}
In our simulation we consider the unconstrained AWGN channel with $m$ channel uses that is described by the vector model $\pmb{y}=\sqrt{\mu_c}\pmb{x}+\pmb{w}$, where $\pmb{x}$ is the transmitted lattice point that is selected randomly from a lattice $\Lambda_c$ with generator matrix $\pmb{G}_c$, $\mu_c$ is the VNR, and $\pmb{w}$ is an AWGN vector with elements that are independent identically distributed, zero-mean Gaussian random variables with unit variance. We consider the Loeliger ensemble of mod-$p$ lattices, where $p$ is a prime. First, we generate the set of all lattices given by $\Lambda_c=\kappa (\mathsf{C}+p\mathbb{Z}^{m})$, where $\kappa$ is a scaling coefficient chosen such that the Voronoi cell volume $V_c=(2\pi e)^{m/2}$, $\mathbb{Z}_p$ denotes the field of mod-$p$ integers, and $\mathsf{C}\subset\mathbb{Z}_p^{m}$ is a linear code over $\mathbb{Z}_p$ with generator matrix in systematic form $[\pmb{I}\;\pmb{P}^\mathsf{T}]^\mathsf{T}$, where $\pmb{P}$ is the parity-check matrix. In the following, we select a lattice code at random with $p=1001$ and fix the code for all simulation results.

\begin{figure}[ht!]
\center
\includegraphics[width=3.5in]{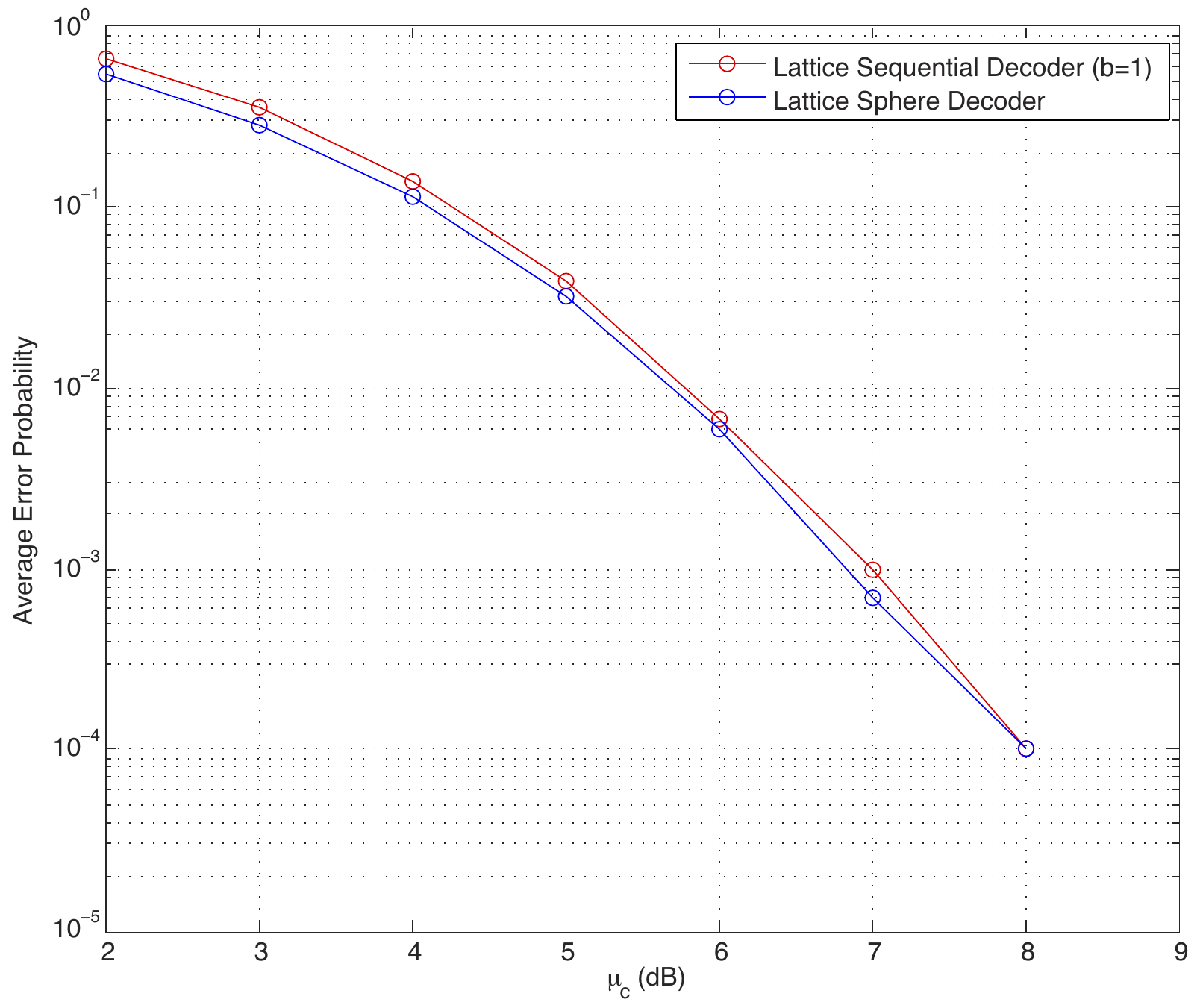}
\caption{Performance comparison between the sphere decoder and the sequential decoder with $b=1$ for a lattice code of dimension $m=30$.}
\label{fig:4}
\end{figure}

\begin{figure}[ht!]
\center
\includegraphics[width=3.5in]{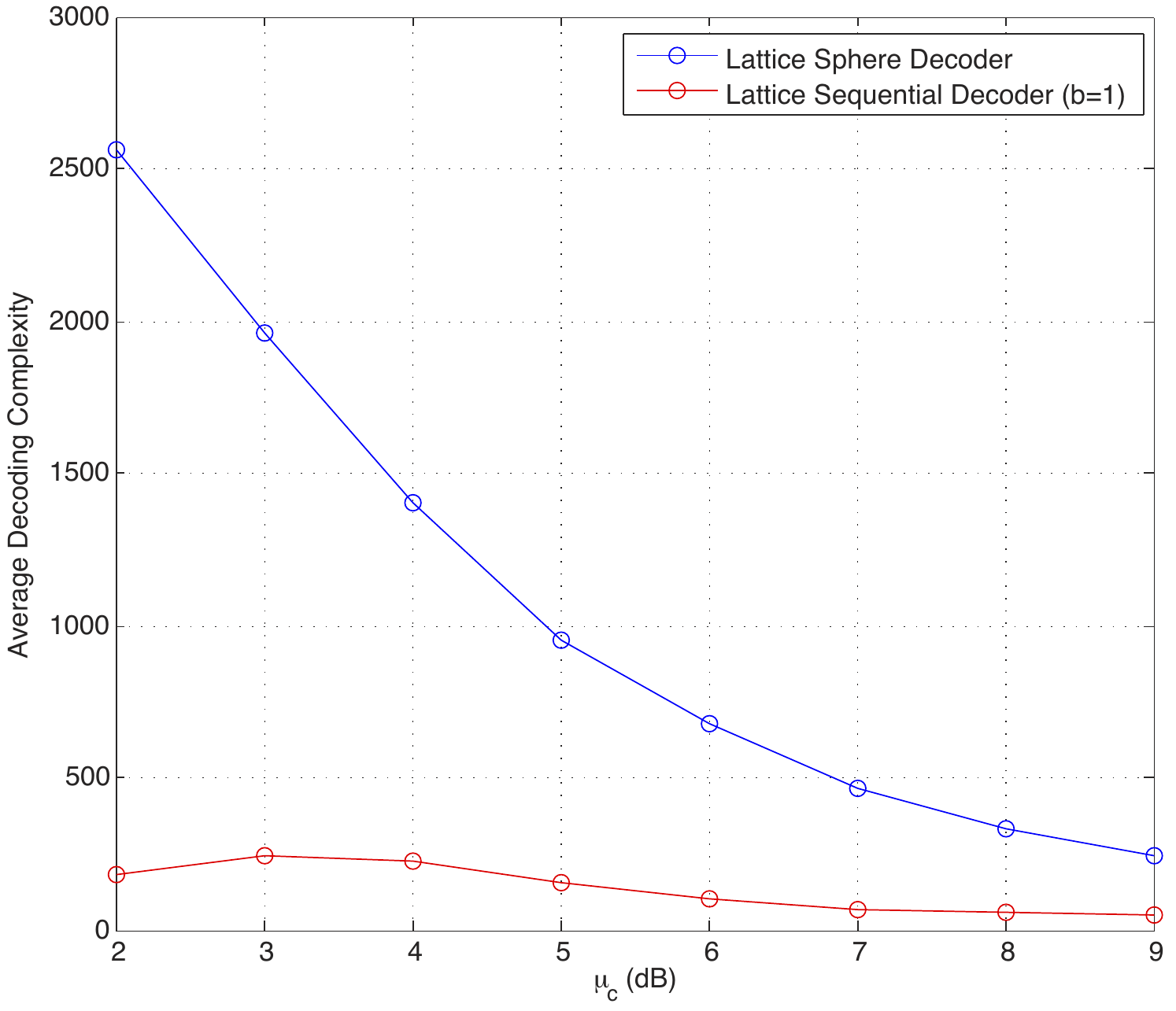}
\caption{Average computational complexity comparison between the lattice sphere decoder and the sequential decoder with bias term $b=1$ for a lattice code of dimension $m=30$.}
\label{fig:5}
\end{figure}

Fig.~\ref{fig:4} and Fig.~\ref{fig:5} demonstrate the great advantage of using the lattice sequential decoder as an alternative to the optimal lattice sphere decoder. The performance and the complexity of both decoders are plotted for a lattice code of length $m=30$. As depicted in Fig.~\ref{fig:5}, there is a significant complexity reduction achieved by using the sequential decoder over the lattice decoder, especially for low-to-moderate VNR, at the expense of very low error performance loss (a fraction of dB), for a bias value $b=1$ (see Fig.~\ref{fig:4}). 

\begin{figure}[ht!]
\center
\includegraphics[width=3.5in]{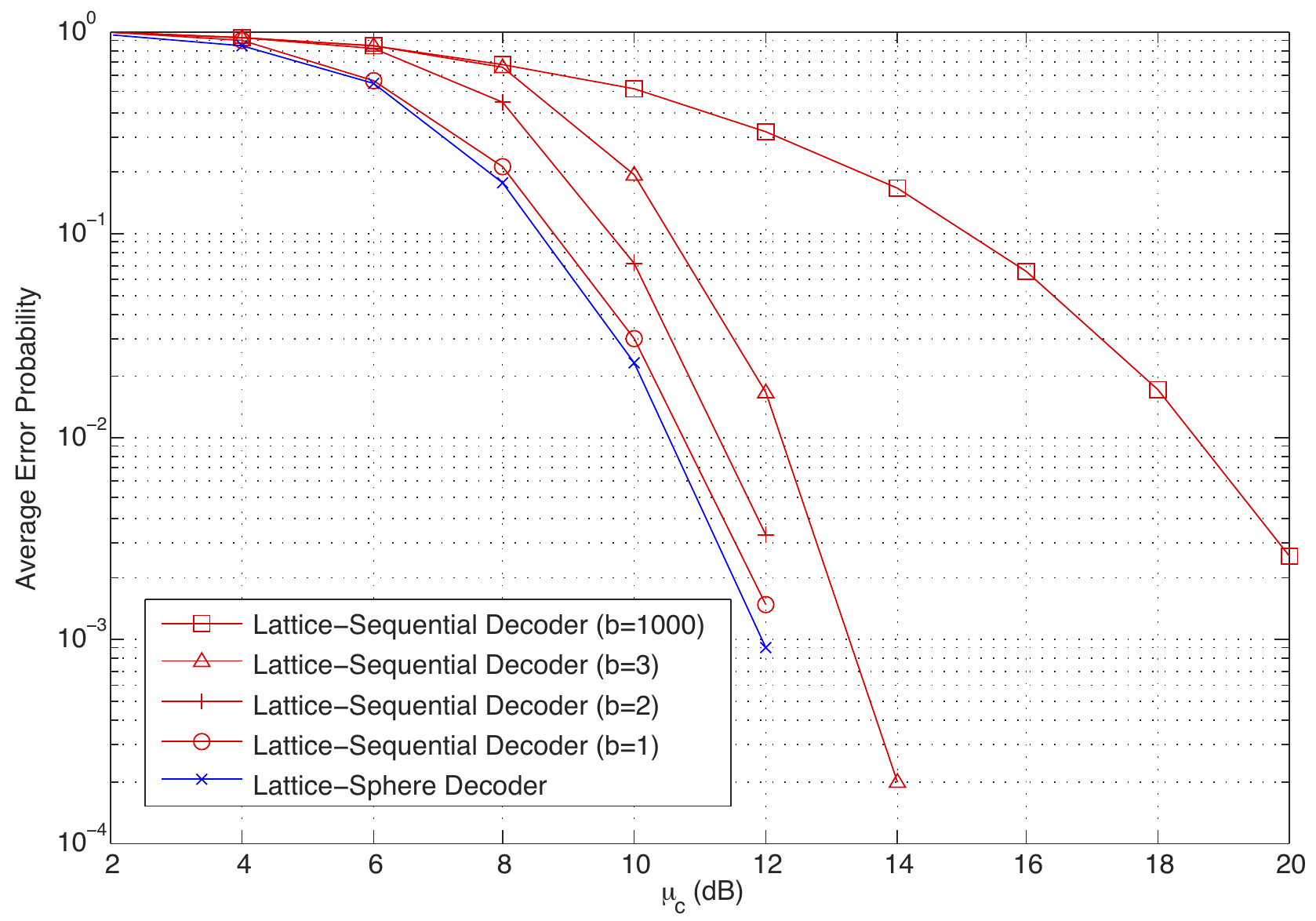}
\caption{Comparison of the lattice sequential decoder's performance for various values of (fixed) bias term.}
\label{fig:6}
\end{figure}

\begin{figure}[ht!]
\center
\includegraphics[width=3.5in]{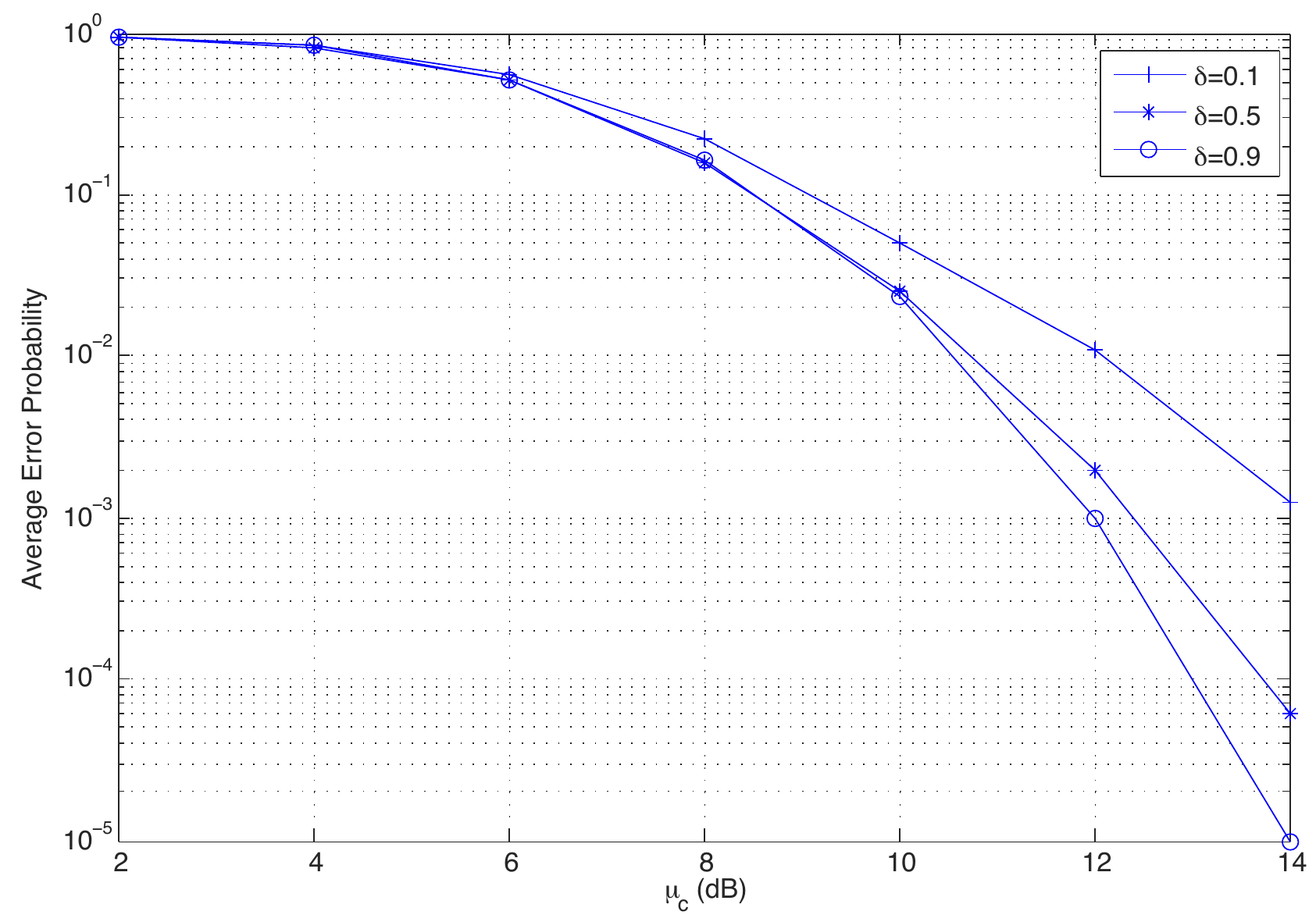}
\caption{Comparison of the lattice sequential decoder's performance when the bias term varies with the VNR as $b_n=(1-\sqrt{\delta})\mu_c$ for several values of $\delta$.}
\label{fig:7}
\end{figure}

Fig.~\ref{fig:6} and Fig.~\ref{fig:7} show the effect of increasing the bias term on the average error probability for the case of fixed and variable bias values, respectively. In Fig.~\ref{fig:6}, we choose fixed bias values (independent of $\mu_c$) and plot the average error probability versus the VNR $\mu_c$ in dB. We also plot the performance of the optimal lattice decoder implemented via the sphere decoder algorithm \cite{DGC} to measure the price of using the sequential decoder in terms of the performance loss. It is clear from the figure that increasing the bias term causes a right-shift to the sequential decoder error probability curve, while maintaining the rate at which the curve decays,\footnote{The asymptotic rate of decay of the error probability curve maybe defined as 
$$\text{slope}\stackrel{\Delta}{=}\lim_{\mu_c\rightarrow \infty}{-\log_e P_e(\mu_c)\over \log_e\mu_c}.$$ Now, for the case of fixed bias, using (\ref{17}) we get 
$$\text{slope}=\lim_{\mu_c\rightarrow \infty}{m\mu_c\over8\log_e\mu_c}-\lim_{\mu_c\rightarrow \infty}{b_n\over4\log_e\mu_c}=\lim_{\mu_c\rightarrow \infty}{m\mu_c\over8\log_e\mu_c},$$ 
which indicates that the slope of the error probability is the same for any finite $b$.} particularly at high VNR values.
\begin{figure}[ht!]
\center
\includegraphics[width=3.55in]{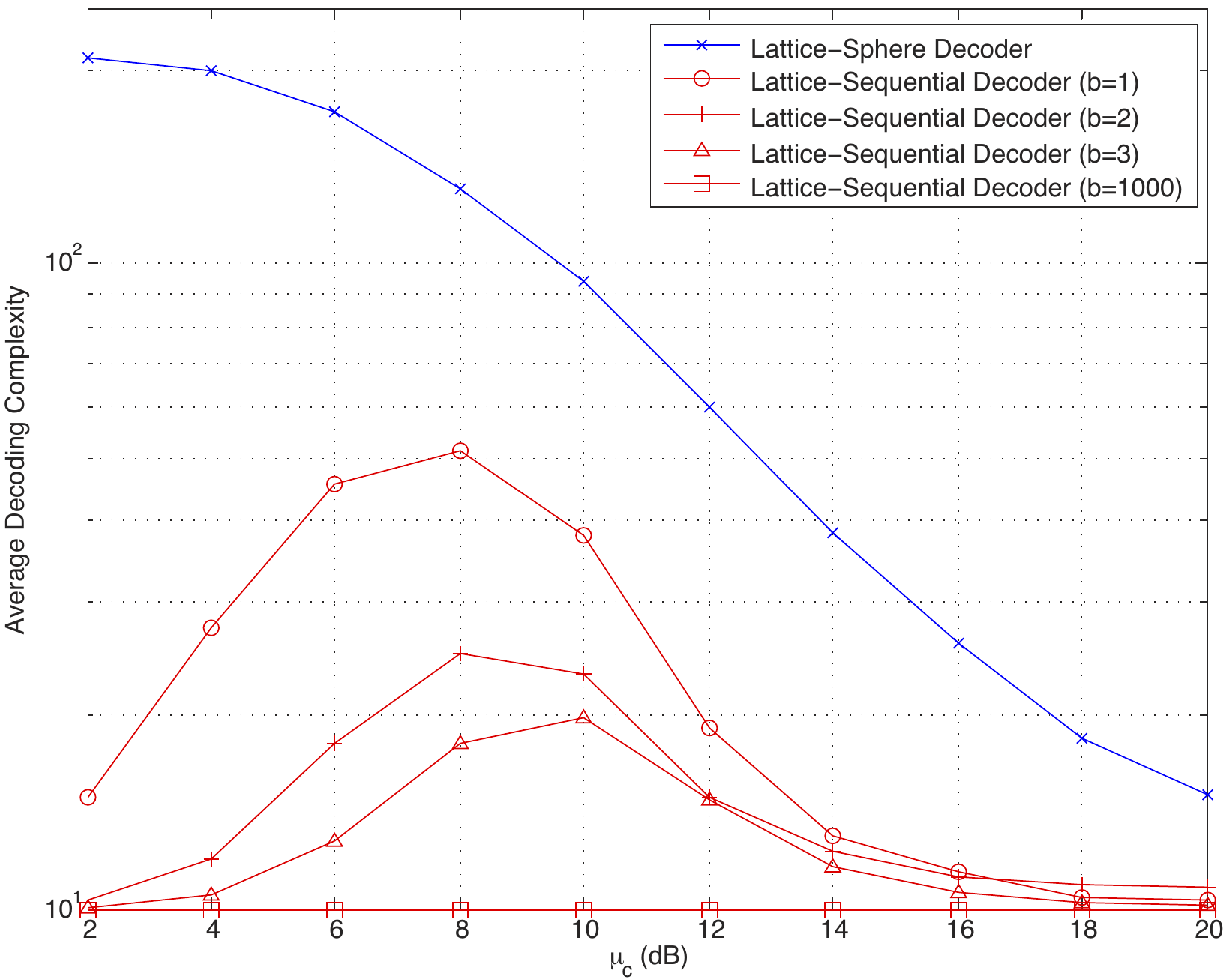}
\caption{The average computational complexity achieved by the sequential decoder for different values of the bias term.}
\label{fig:8}
\end{figure}
This basically agrees with the derived theoretical results provided in (\ref{exp_approx}) and (\ref{17}). On the other hand, if we let $b$ to scale linearly with VNR as $(1-\sqrt{\delta})\mu_c$, where $0\leq\delta\leq 1$ (see (\ref{188})), then according to the error exponent analysis, we expect that the rate of decay (slope) of the error probability curve would decrease\footnote{In this case, the rate of decay (slope) of the error probability curve can be shown to be equal to $\delta [m\mu_c/8\log_e\mu_c]$ which depends on $b$ via $\delta$.} as we decrease $\delta$. This is depicted in Fig.~\ref{fig:7}, which also agrees with the derived theoretical results.

Finally, Fig.~\ref{fig:8} shows the effect of increasing the bias term on the average computational complexity (defined as the total number of visited nodes during the search). For comparison, we also include in Fig.~\ref{fig:8} the average complexity of the sphere decoder for the same lattice code. The average complexity is plotted versus the VNR in dB. It is clear that for all values of $b$ the sequential decoder has much lower complexity compared to the lattice (sphere) decoder, especially for low-to-moderate VNRs. The reason for the bell-like shape of the average complexity that occurs at low-to-moderate VNRs is due to the fact that with high-probability the received signal is close to the edge of the Voronoi cell. This basically requires the decoder to visit more nodes in the tree before decoding the message. As the VNR decreases or increases, the received signal becomes closer to a wrong lattice point or to the transmitted lattice point, respectively, which allows the decoder to decode the message without visiting many nodes. This leads to the very low average complexity as depicted in Fig.~\ref{fig:8}. The result also shows that as we increase the bias term, the average complexity significantly reduces, especially for low-to-moderate VNR values. As $b\rightarrow\infty$, the number of computations becomes equal to $m$ (the signal dimension) for all VNR values. This agrees with the derived theoretical results.

In conclusion, simulation results indicate that increasing the bias term in the decoding algorithm significantly reduces the complexity but at the expense of losing performance.

\section{Conclusion}
In this paper, we have analyzed the performance limits and the computational complexity of the lattice sequential decoder applied to the unconstrained AWGN channel. The error probability has been analyzed following the footsteps of Poltyrev by deriving the error exponent of the sequential decoder as a function of the VNR and the decoding parameter---the bias term. The bias term is responsible for the performance-complexity tradeoff achieved by the decoder. It has been shown (analytically and via simulation) that, if the bias term is fixed and independent of the VNR, then increasing the bias term causes only a right-shift to the error probability curve (occurs as a loss in the coding gain). However, if the bias term is scaled linearly with the VNR, the rate at which the error probability curve decays gets affected accordingly. It has also been shown that increasing the bias term significantly reduces the average number of computations required by the decoder to decode a message. However, the price of the complexity improvements comes at the expense of a loss in the performance. Hence, a fundamental trade-off exists between the error performance, the decoding complexity, and the VNR. 

By revealing the tradeoff between performance, complexity and lattice density, it introduces the concept of lattice density into lattice decoding for the first time making it a promising area in 
lattice applications for digital and wireless communications. An interesting venue for future work is to derive bounds on the moments 
of sequential decoding complexity. As shown in this paper, since the decoding complexity is random, there exists a non-zero probability that the amount of computations performed by the decoder 
may become excessive causing a buffer overflow which is considered an 
important metric for the design of a sequential decoder. Therefore, studying these moments (e.g., the variance of the decoding complexity) 
is important to obtain estimates on the probability of buffer overflow \cite{Savage}.

\appendices


\section*{Acknowledgment}
The authors would like to thank the Editor for his diligence and the reviewers for their detailed comments.

\ifCLASSOPTIONcaptionsoff
  \newpage
\fi




\begin{thebibliography}{99}
\bibitem{Conway} J.~H.~Conway and N.~J.~A.~Sloane, \textit{Sphere Packings, Lattices, and Groups}, 3rd ed. Springer Verlag NewYork, 1999.

\bibitem{deBuda} R.~deBuda, ``The upper bound of a new near-optimal code,'' \textit{IEEE Trans. on Inform. Theory}, vol.~IT-21, no.~7 pp.~441-445, July 1975.

\bibitem{Polytrev}G.~ Poltyrev, ``On coding without restrictions for the AWGN channel,'' \textit{IEEE Trans.~Inform.~Theory}, vol.~40, no.~2, pp.~409-417, Mar. 1994.

\bibitem{Loe} H.~Loeliger, ``Averaging bounds for lattices and linear codes,'' \textit{IEEE Trans.~Inform.~Theory}, vol.~43, no.~6, pp.~1767-1773, Nov.~1997.

\bibitem{Ubranke_Rimoldi} R.~Urbanke and B. Rimoldi, ``Lattice codes can achieve capacity on the AWGN channel,'' \textit{IEEE Trans.~on Inform.~Theory}, vol.~44, no.~1, pp.~273-278, Jan. 1998.

\bibitem{AZ}  A.~Ingber and R.~Zamir, ``Expurgated infinite constellations at finite dimensions,'' in \textit{Proc. IEEE Int'l Symp. Inform. Theory}, (ISIT'12) MA, USA, July~2012.

\bibitem{CLSP} E.~Agrell, T.~Eriksson, A.~Vardy, and K.~Zeger, ``Closest point search in lattices,'' \textit{IEEE Trans.~on Inform.~Theory}, vol.~48, no.~8, pp.~2201Ð2214, Aug.~2002.

\bibitem{EZ2} U.~Erez, S.~Litsyn, and R.~Zamir, ``Lattices which are good for (almost) everything,'' \textit{IEEE Trans.~Inform.~Theory}, vol.~51, no.~10, pp.~3401-3416, Oct.~2005.

\bibitem{Li} L.-C. Choo, C. Ling, and K.-K. Wong, ``Achievable rates for lattice coding over the Gaussian wiretap channel," in \textit{Proc. IEEE Physical Layer Security Workshop in Conjunction with IEEE Int'l Comm. Conf. (ICC'11)}, Kyoto, Japan, June 2011. 

\bibitem{Hlawka} H. Minkowski, ``Zur Geometrie der Zahlen,'' \textit{Math.~Z.}, vol.~49, pp.~285-312, 1944.

\bibitem{Rogers} C.~A.~Rogers, \textit{Packing and Covering}, Cambridge, UK: Cambridge Uni.~Press, 1964.

\bibitem{DGC} B.~Hassibi and H.~Vikalo, ``On sphere decoding algorithm. Part I: expected complexity,'' \textit{IEEE Trans. Sign. Proc.}, vol.~53, no. 8, pp.~2389-2401, Aug.~2005.

\bibitem{JB} J.~Boutros, N.~Gresset, L.~Brunel, and M.~Fossorier, ``Soft-input soft-output lattice sphere decoder for linear channels,'' in \textit{Proc. IEEE Global Comm. Conf. (GLOBECOMÕ03)}, San Francisco, USA, Dec.~2003.

\bibitem{Jacobs} I.~M.~Jacobs and E.~R.~Berlekamp, ``A lower bound to the distribution of computation for sequential decoding,'' \textit{IEEE Trans.~Inform.~Theory}, vol.~IT-13, pp.~167-174, April 1976.

\bibitem{Stack} F. ~Jelinek, ``A fast sequential decoding algorithm using a stack,'' \textit{IBM J.~Res.~Dev.~}, vol.~13, pp.~675-685, Nov.1969.

\bibitem{MGDC} A. Murugan, H. El Gamal, M. O. Damen, and G. Caire, ``A unified framework for tree search decoding: Rediscovering the sequential decoder,'' \textit{IEEE Trans. Inform.~ Theory}, vol.~52, no.~3, pp.~933-953, Mar. 2005.


\bibitem{TVZ} V. Tarokh, A. Vardy, and K. Zeger, ``Sequential decoding of lattice codes,'' in \textit{Proc. IEEE Int'l Symp. Inform. Theory}, (ISIT'97), Ulm, Germany, June 1997.

\bibitem{OMN} O. Shalvi, N.~Sommer, and M.~Feder, ``Signal codes: Convolutional lattice codes,'' \textit{IEEE Trans.~on Inform.~Theory}, vol.~57, no. 8, pp.~5203-5226, Aug. 2011.

\bibitem{Forney2} G. D. Forney Jr., M. D. Trott, and S. Chung, ``Sphere-bound-achieving coset codes and multilevel coset codes,'' \textit{IEEE Trans.~on Inform.~Theory}, vol.~46, no. 3, pp.~820-850, May 2000.     
    
\bibitem{Gallager} R. Gallager, \textit{Information Theory and Reliable Communication}. New
York: John Wiley and Sons, 1968. 

\bibitem{Savage} J.~E.~Savage, ``The computation problem with sequential decoding'', M.~I.~T Lincoln Lab., Lexington, Mass., Tech. Rept. 371, Feb. 1965.
    


\end{thebibliography}
%

%

  \begin{IEEEbiography}[{\includegraphics[width=1in,height=1.25in,clip,keepaspectratio]{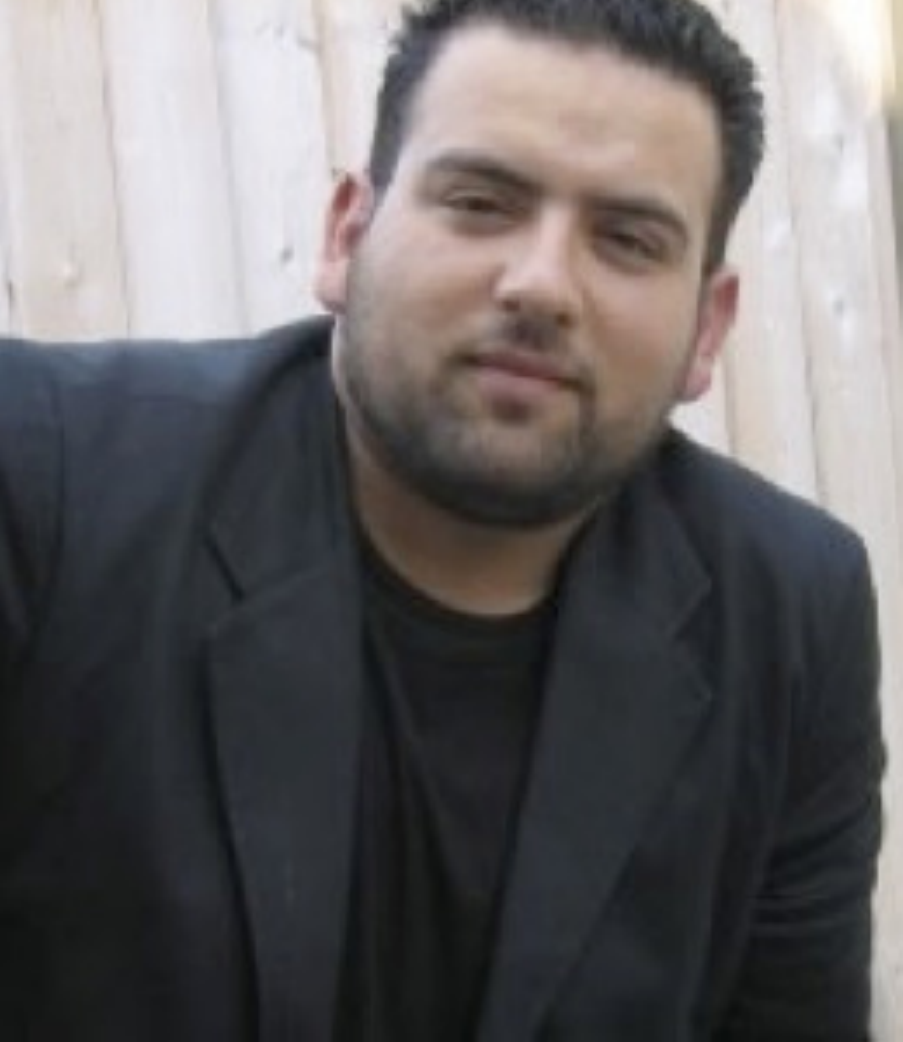}}]{Walid Abediseid} (S'04, M'11) was
born in Etobicoke, Ontario, Canada. He received the B.Sc. and M.Sc. degrees in Electrical Engineering from the University of Ottawa, Canada, in 2004 and 2007, respectively. He then received his Ph.D. from the Department of Electrical and Computer Engineering, University of Waterloo in 2011. He is a postdoctoral fellow at King Abdullah University of
Science and Technology (KAUST), Thuwal, Makkah Province, Saudi
Arabia, since December 2011. His research interests include coding and information theory, MIMO wireless communication systems, lattice applications for digital and wireless communications, detection and estimation.

Dr. Abediseid was a recipient of Research In Motion Graduate Scholarship from 2008 to 2009, and the NSERC Alexander Graham Bell Canada Graduate Scholarship --- one of Canada's most prestigious graduate 
awards from 2009 to 2011.
\end{IEEEbiography}

  \begin{IEEEbiography}[{\includegraphics[width=1in,height=1.25in,clip,keepaspectratio]{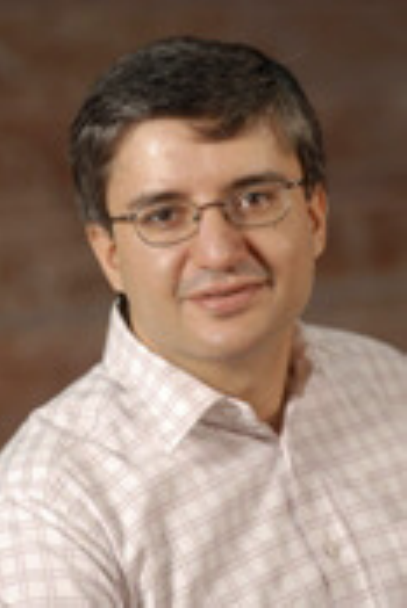}}]{Mohamed-Slim Alouini} (S'94, M'98, SM'03, F’09) was
born in Tunis, Tunisia. He received the Ph.D. degree in Electrical Engineering
from the California Institute of Technology (Caltech), Pasadena,
CA, USA, in 1998. He served as a faculty member in the University of Minnesota,
Minneapolis, MN, USA, then in the Texas A\&M University at Qatar,
Education City, Doha, Qatar before joining King Abdullah University of
Science and Technology (KAUST), Thuwal, Makkah Province, Saudi
Arabia as a Professor of Electrical Engineering in 2009. His current
research interests include the modeling, design, and
performance analysis of wireless communication systems.
\end{IEEEbiography}




\end{document}